# Conformable Fractional Polytropic Gas Spheres

E. A.-B. Abd-Elsalam[1] and Mohamed I. Nouh[2]

[1]Department of Mathematics, Faculty of Science, New Valley University, El-Kharja 72511, Egypt.

[2]Department of Astronomy, National Research Institute of Astronomy and Geophysics, Helwan, Cairo 17211, Egypt.

**Abstract:** Lane –Emden differential equation of the polytropic gas sphere could be used to construct simple models of stellar structures, star clusters and many configurations in astrophysics. This differential equation suffers from the singularity at the center and has an exact solution only for the polytropic index $n=0, 1$ and $5$. In the present paper, we present an analytical solution to the fractional polytropic gas sphere via accelerated series expansion. The solution is performed in the frame of conformable fractional derivatives. The calculated models recover the well-known series of solutions when $\alpha = 1$. Physical parameters such as mass-radius relation, density ratio, pressure ratio and temperature ratio for different fractional models have been calculated and investigated. We found that the present models of the conformable fractional stars have smaller volume and mass than that of both the integer star and fractional models performed in the frame of modified Rienmann Liouville derivatives.

## 1. Introduction

In last decades, fractional calculus has important applications in physics and engineering, such as particle physics, wave mechanics, electrical systems, fractal wave propagations, and various real-life problems are described exactly based on the fractional differential equations, Herrmann (2014).

Lane–Emden equation has a long history in modeling several phenomena in mathematical physics and astrophysics such as the theory of stellar structure, the thermal behavior of a spherical cloud of gas, isothermal gas spheres, and the theory of thermionic currents. The main difficulty of solving this Lane–Emden equation comes from the singular behavior that occurs at the center of the polytrope. Many authors introduced several approaches to solving the integer version of the Lane-Emden equation, Chowdhury (2009), Ibrahim and Darus (2008), Momani and Ibrahim (2008), Nouh (2004), Hunter (2001).

The fractional version of the Newtonian stellar polytrope has been investigated by Bayian and Krisch (2015) for the incompressible gas sphere, Abdel-Salam and Nouh (2016) introduced a series solution for the fractional isothermal gas sphere and obtained a solution that converges to the surface of the sphere with only a few series terms compared with the solution of the integer differential equation and Nouh and Abd-Elsalam (2018) for the general form of the Lane-Emden equation.

The polytropic Lane-Emden equation in its fractional form is given by (Nouh and Abd-Elsalam, 2018)

$$\frac{1}{x^{2\alpha}} \frac{d^\alpha}{dx^\alpha}\left( x^{2\alpha} \frac{d^\alpha u}{dx^\alpha} \right) = -u^n. \qquad (1)$$

Where the dimensionless function $u$ (Emden function) is given by

$$\rho = \rho_c u^n,$$

$\rho$ and $\rho_c$ are the density and central density respectively, and $x$ is a dimensionless variable.

In the present article, we introduce an approximate solution for the conformable fractional polytropic gas sphere. We construct a recurrence relation for the coefficient of the power series. The physical parameters of the polytrope are deduced. The structure of the paper is as follows. In section 2, principles of the conformable fractional derivatives are introduced. The series solution to the fractional polytropic gas sphere is described in section 3. The present results are presented in section 4 and section 5 is devoted to the conclusion reached.

## 2. Conformable Fractional Derivatives

There are various definitions of the fractional derivatives. Examples include Riemann–Liouville, Caputo, modified Riemann–Liouville, Kolwankar–Gangal, Cresson's and Chen's fractal derivatives, Mainardi (2010) and Herrmann (2014).

Khalil et al (2014) introduced the conformable fractional derivative by using the limits in the form

$$D^\alpha f(t) = \lim_{\varepsilon \to 0} \frac{f(t+\varepsilon t^{1-\alpha}) - f(t)}{\varepsilon} \quad \forall t > 0, \, \alpha \in (0,1], \tag{2}$$

$$f^{(\alpha)}(0) = \lim_{t \to 0^+} f^{(\alpha)}(t). \tag{3}$$

Here $f^{(\alpha)}(0)$ is not defined. When $\alpha = 1$, this fractional derivative reduces to the ordinary derivative. The conformable fractional derivative has the following properties:

$$D^\alpha t^p = p t^{p-\alpha}, \quad p \in \mathbb{R}, \quad D^\alpha c = 0, \quad \forall f(t) = c, \tag{4}$$

$$D^\alpha (a f + b g) = a D^\alpha f + b D^\alpha g, \quad \forall a, b \in \mathbb{R}, \tag{5}$$

$$D^\alpha (f g) = f D^\alpha g + f D^\alpha g, \tag{6}$$

$$D^\alpha f(g) = \frac{df}{dg} D^\alpha g, \quad D^\alpha f(t) = t^{1-\alpha} \frac{df}{dg}, \tag{7}$$

where $f$, $g$ are two $\alpha$-differentiable functions and $c$ is an arbitrary constant. Equations (5) to (6) are proved by Khalil et al in [28]. The conformable fractional derivative of some functions

$$D^\alpha e^{ct} = c t^{1-\alpha} e^{ct}, \quad D^\alpha \sin(ct) = c t^{1-\alpha} \cos(ct), \quad D^\alpha \cos(ct) = -c t^{1-\alpha} \sin(ct),$$
$$D^\alpha e^{ct^\alpha} = c \alpha e^{ct^\alpha}, \quad D^\alpha \sin(ct^\alpha) = c \alpha \cos(ct^\alpha), \quad D^\alpha \cos(ct^\alpha) = -c \alpha \sin(ct^\alpha). \tag{8}$$

## 3. Series Solution to the Fractional Polytropic Gas Spheres

### 3.1. Successive Fractional Derivatives of the Emden Function

The fractional Lane-Emden equation (Equation 1) could be written as

$$x^{-2\alpha} D_x^\alpha \left( x^{2\alpha} D_x^\alpha \right) u + u^n = 0, \tag{9}$$

with the initial conditions

$$u(0) = 1, \quad D_x^\alpha u(0) = 0 \tag{10}$$

where $u = u(x)$, is the Emden function and $0 < \alpha \leq 1$.

By assuming the transform $X = x^\alpha$, the solution could be expressed in a series form as

$$u(X) = \sum_{m=0}^{\infty} A_m X^m, \tag{11}$$

Using the first initial condition (Equation (10)) to Equation (11) we get $A_0 = 1$ and applying Equations (4) and (5) to Equation (11) we get

$$D_x^\alpha u = \alpha A_1 + 2\alpha A_1 x^\alpha + 3\alpha A_2 x^{2\alpha} + \ldots \qquad (12)$$

Applying the second initial condition we obtain

$$D_x^\alpha u(0) = \alpha A_1$$

Then, $A_1 = 0$ and the series expansion could be written as

$$u(X) = 1 + \sum_{m=2}^{\infty} A_m X^m \qquad (13)$$

Apply the second derivative of the Emden function $u$, we get

$$D_x^{\alpha\alpha} u = D_x^\alpha D_x^\alpha u = 2\alpha^2 A_2 + 6\alpha^2 A_3 x^\alpha + 12\alpha^2 A_3 x^{2\alpha} + \ldots \qquad (14)$$

At $X = 0$ we have

$$D_x^{\alpha\alpha} u(0) = 2\alpha^2 A_2,$$

Applying $j$ times derivatives to the last equation we obtain

$$D^{\overbrace{\alpha \ldots \alpha}^{j\text{ times}}} u(0) = D_x^\alpha \ldots D_x^\alpha u(0) = j! \alpha^j A_j,$$

where $A_j$ are constants to be determined.

Now suppose that

$$u^n = G(X) = \sum_{m=0}^{\infty} Q_m X^m, \qquad (15)$$

At $X = 0$ we have after $j$ times derivatives

$$D_x^\alpha \ldots D_x^\alpha G(0) = j! \alpha^j Q_j,.$$

### 3.2. Fractional Derivative of Emden Function Raised to Powers

To obtain the fractional derivative of the Emden function $u^n$, we apply the fractional derivative of the product of two functions $u^2$, it will be considered as $u$ times $u$. Similarly, $u^3$ will be considered as $u$ times $u^2$ and so on. Taking the fractional derivative for both sides of Equation (14), we have

$$D_x^\alpha u^n = D_x^\alpha G,$$

that is

$$nu^{n-1}D_x^\alpha u = D_x^\alpha G .$$

or

$$n u^n D_x^\alpha u = u D_x^\alpha G . \qquad (16)$$

Differentiating both sides of Equation (15) $k$ times we have

$$D^{\overbrace{\alpha \cdots \alpha}^{k\text{ times}}} [nGD_x^\alpha u] = D^{\overbrace{\alpha \cdots \alpha}^{k\text{ times}}} (u D_x^\alpha G) ,$$

Then we have

$$n\sum_{j=0}^{k} \binom{k}{j} D^{\overbrace{\alpha \cdots \alpha}^{j+1\text{ times}}} u D^{\overbrace{\alpha \cdots \alpha}^{k-j\text{ times}}} G = \sum_{j=0}^{k} \binom{k}{j} D^{\overbrace{\alpha \cdots \alpha}^{j+1\text{ times}}} G D^{\overbrace{\alpha \cdots \alpha}^{k-j\text{ times}}} u,$$

at $X=0$, we have

$$n\sum_{j=0}^{k} \binom{k}{j} D^{\overbrace{\alpha \cdots \alpha}^{j+1\text{ times}}} u(0) D^{\overbrace{\alpha \cdots \alpha}^{k-j\text{ times}}} G(0) = \sum_{j=0}^{k} \binom{k}{j} D^{\overbrace{\alpha \cdots \alpha}^{j+1\text{ times}}} G(0) D^{\overbrace{\alpha \cdots \alpha}^{k-j\text{ times}}} u(0),$$

or

$$n\sum_{j=0}^{k} \binom{k}{j}(j+1)!\alpha^{(j+1)}A_{j+1}(k-j)!\alpha^{(k-j)}Q_{k-j} = \sum_{j=0}^{k} \binom{k}{j}(j+1)!\alpha^{(j+1)}Q_{j+1}(k-j)!\alpha^{(k-j)}A_{k-j}, \Rightarrow$$

$$n\sum_{j=0}^{k} \binom{k}{j}(j+1)!(k-j)!\alpha^{(k+1)}A_{j+1}Q_{k-j} = \sum_{j=0}^{k} \binom{k}{j}(j+1)!(k-j)!\alpha^{(k+1)}Q_{j+1}A_{k-j},$$

So, we get the following equations

$$n\sum_{j=0}^{k} \frac{k!(j+1)!(k-j)!\alpha^{(k+1)}}{j!(k-j)!} A_{j+1}Q_{k-j} = \sum_{j=0}^{k} \frac{k!(j+1)!(k-j)!\alpha^{(k+1)}}{j!(k-j)!} A_{k-j}Q_{j+1}$$

$$n\sum_{j=0}^{k} k!(j+1)\alpha^{(k+1)} A_{j+1}Q_{k-j} = \sum_{j=0}^{k} k!(j+1)\alpha^{(k+1)} A_{k-j}Q_{j+1},$$

$$n\sum_{j=0}^{k} k!(j+1) A_{j+1}Q_{k-j} = \sum_{j=0}^{k} k!(j+1) A_{k-j}Q_{j+1},$$

$$n\sum_{j=0}^{k} k!(j+1) A_{j+1}Q_{k-j} = \sum_{j=0}^{k-1} k!(j+1) A_{k-j}Q_{j+1} + k!(k+1)A_0 Q_{k+1},$$

and

$$k!(k+1)A_0Q_{k+1} = n\sum_{j=0}^{k} k!(j+1)A_{j+1}Q_{k-j} - \sum_{j=0}^{k-1} k!(j+1)A_{k-j}Q_{j+1}.$$

In the last equation, let $i = j + 1$ in the first summation and $i = k - j$ in the second summation, we get

$$k!(k+1)A_0Q_{k+1} = n\sum_{i=1}^{k+1} k!(i)A_iQ_{k+1-i} - \sum_{i=1}^{k} k!(k+1-i)A_iQ_{k+1-i},$$

if $m = k+1$, then

$$m!A_0Q_m = n\sum_{i=1}^{m} (m-1)!(i)A_iQ_{m-i} - \sum_{i=1}^{m-1} (m-1)!(m-i)A_iQ_{m-i},$$

By adding the zero value $\{-(m-1)!(m-m)A_mQ_0\}$ to the second summation, we get

$$m!A_0Q_m = n\sum_{i=1}^{m} (m-1)!(i)A_iQ_{m-i} - \sum_{i=1}^{m-1} (m-1)!(m-i)A_iQ_{m-i} - (m-1)!(m-m)A_mQ_0,$$

$$m!A_0Q_m = n\sum_{i=1}^{m} (m-1)!(i)A_iQ_{m-i} - \sum_{i=1}^{m} (m-1)!(m-i)A_iQ_{m-i},$$

then the coefficients $Q_m$ could be written as

$$Q_m = \frac{1}{m!A_0}\sum_{i=1}^{m} (m-1)!(in-m+i)A_iQ_{m-i}, \ \forall \ m \geq 1, \tag{17}$$

where

$$A_0 = 1, \ A_1 = 0, \quad Q_0 = A_0^n = 1, \ Q_1 = \frac{n}{A_0}A_1Q_0 = 0.$$

**3.3. Fractional Derivatives of the Series Expansion of the Emden Function**

Using $u = 1 + \sum_{m=2}^{\infty} A_m X^m$ we obtain

$$D_x^\alpha u = \sum_{m=2}^{\infty} \alpha A_m mX^{m-1}x^{\alpha-\alpha} = \sum_{m=2}^{\infty} \alpha A_m mX^{m-1}, \tag{18}$$

The second derivative of the Emden function $u$ could be given by

$$D_x^\alpha D_x^\alpha u = \sum_{m=2}^{\infty} \alpha^2 A_m m(m-1)X^{m-2}x^{\alpha-\alpha} = \sum_{m=2}^{\infty} \alpha^2 A_m m(m-1)X^{m-2},, \tag{19}$$

Substituting Equations (18) and (19) in Equation (9) we have

$$x^{-2\alpha}D_x^{\alpha}\left(x^{2\alpha}D_x^{\alpha}\right)u+u^n=0, \Rightarrow x^{-2\alpha}[x^{2\alpha}D_x^{\alpha\alpha}u+2\alpha x^{\alpha}D_x^{\alpha}u]+u^n=0,$$

$$D_x^{\alpha\alpha}u+2\alpha x^{-\alpha}D_x^{\alpha}u+u^n=0,$$

$$x^{2\alpha}\sum_{m=2}^{\infty}\alpha^2 A_m m(m-1)X^{m-2}+2\alpha x^{\alpha}\sum_{m=2}^{\infty}\alpha A_m mX^{m-1}+x^{2\alpha}\left[1+\sum_{m=2}^{\infty}Q_m X^m\right]=0,$$

or

$$\sum_{m=2}^{\infty}\alpha^2 A_m m(m-1)X^m+\sum_{m=2}^{\infty}2\alpha^2 A_m mX^m+X^2+\sum_{m=2}^{\infty}Q_m X^{m+2}=0,$$

$$\sum_{m=2}^{\infty}\alpha^2 A_m m(m-1+2)X^m+X^2+\sum_{m=2}^{\infty}Q_m X^{m+2}=0, \tag{20}$$

$$\sum_{m=2}^{\infty}\alpha^2 A_m m(m+1)X^m+X^2+\sum_{m=2}^{\infty}Q_m X^{m+2}=0.$$

By putting $m=k+2$ in the first part and $m=k$ in the third part of Equation (19), we get

$$\sum_{k=0}^{\infty}\alpha^2 A_{k+2}(k+2)(k+3)X^{k+2}+X^2+\sum_{k=2}^{\infty}Q_k X^{k+2}=0,$$

After some manipulations, we get the recurrence relation of the coefficients as

$$\alpha^2(k+2)(k+3)A_{k+2}+Q_k=0, \quad \forall\ k\geq 2.$$

### 3.4. Recurrence Relations

Now we can determine the coefficients of the series expansion from the following two recurrence relations

$$A_{k+2}=-\frac{Q_k}{\alpha^2(k+2)(k+3)}, \quad \forall\ k\geq 2 \tag{21}$$

and

$$Q_m=\frac{1}{m!A_0}\sum_{i=1}^{m}(m-1)!(in-m+i)A_i Q_{m-i}, \quad \forall\ m\geq 1, \tag{22}$$

If we put $\alpha=1$ in Equations (21) and (22) the series coefficients are reduced to the well-known series solution of the integer LEE. For $k=0,\ 1,\ 2,\ 3$ in Equations (21) and (22) we get

$$A_0=1,\ A_1=0,\quad Q_0=1,\ Q_1=0,\ A_2=-\frac{1}{6\alpha^2},\ A_3=0,\ A_4=\frac{n}{120\alpha^4},\ A_5=0.$$

Then the series solution at $\alpha = 1$ is reduced to the well-known solution of Equation (9) as

$$u_n(x) = 1 - \frac{1}{6}x + \frac{n}{120}x^2 - \ldots$$

## 4. Results

The analytical solution of Equation (9) with the initial conditions in Equation (10) determines the polytropic structure of the configuration. This solution is represented by Equation (13) together with Equations (21) to (22). First, we compare the present solution with the solution presented by Nouh and Abdel-Salam (2018), where they have simulated the fractional polytropic gas sphere via accelerated series expansion; the results indicated that the fractional polytropic sphere is smaller than the integer one. We updated our code to the conformable formulations, so the modified code contains the two series expansions. We run the code for the polytrope with n=3 at various values of $\alpha$.

The results are listed in Table 1, where column 1 represents the results from the solution by Nouh and Abdel-Salam (2018); hereafter we shall call it FP; and column 2 represents the results from the present algorithm (hereafter we shall call it CFP), we fixed the number of series terms for all calculations. The radius of convergence $x_1$ of the power series solution without applying acceleration techniques is limited.

To improve the radius of convergence of the divergent series, we implemented the scheme proposed by Nouh (2004) to enable the series to reach the surface of the polytrope, the results are listed in the second row of Table 2. The situation is different than that appeared in Table 1; the CFP has a radius less than that of the FP, which consequently means that the CFP polytrope has a lower volume than FP polytrope and the integer one.

Table 1: Radius of convergence of the n=3 fractional polytrope without series acceleration.

| $\alpha$ | $X_{1\ FP}$ | $X_{1\ CFP}$ |
|---|---|---|
| 1 | 2.46 | 2.46 |
| 0.99 | 2.37 | 2.45 |
| 0.98 | 2.32 | 2.44 |
| 0.97 | 2.19 | 2.435 |
| 0.96 | 2.1 | 2.43 |
| 0.95 | 2 | 2.42 |

The mass contained in a radius $r$, radius, pressure, and temperature of the polytrope are given by (Nouh and Abd-Elsalam, 2018)

$$M(x^\alpha) = 4\pi \left[\frac{K(n+1)}{4\pi G}\right]^{\frac{3}{2}} \rho_c^{\frac{3-n}{2n}} \left[-\left(x^{2\alpha} \frac{d^\alpha u}{d x^\alpha}\right)\right]_{x=x_1},$$

$$R^\alpha = \left[\frac{K(n+1)}{4\pi G}\right]^{\frac{1}{2}} \rho_c^{\frac{1-n}{2n}} x_1^\alpha,$$

$$P = P_c\, u^{n+1}, \tag{47}$$

and

$$T = T_c\, u^n. \tag{48}$$

The central density is computed from the equation

$$\rho_c = -\frac{x^{2\alpha} M_0}{4\pi R_0^3 \left[\left(\frac{d^\alpha u}{d x^\alpha}\right)\right]_{x=x1}} \tag{49}$$

We first calculate a fractional model with a polytropic index n=3 suitable for the sun and solar-type stars. To calculate the solar physical parameters, we take the mass, radius and central temperature of the sun as, $M_0 = 1.989 \times 10^{33}$ gm, $R_0 = 6.9598 \times 10^8$ m and $T_c = 1.570 \times 10^7$ K respectively. The results are tabulated in Table 2, where row 1 represents the fractional parameter $\alpha$, row 2 is the first zero $X_{1\ FP}$ calculated by Nouh and Abdel-Salam (2018), row 3 is the first zero $X_{1\ CFP}$ calculated by the present algorithm, row 4 is the radius of the polytrope calculated by Nouh and Abdel-Salam (2018), row 5 is the radius of the polytrope calculated by the present algorithm, row 6 is the mass calculated by Nouh and Abdel-Salam (2018) and row 7 is the mass calculated by the present algorithm. The results indicate that the CFP star has smaller volume and mass than the FP and integer stars. Figures 1, 2 and 3 illustrate the distribution of the Emden function, pressure distribution and mass-radius relation for the polytrope with n=3. The effects of the fractional parameter on those distributions is remarkable, especially for the mass-radius relation. We note that the CFP spheres have a radius smaller than that of the FP spheres.

Table 2: Accelerated fractional polytropic gas spheres with n=3.

| $\alpha$ | 1 | 0.99 | 0.98 | 0.97 | 0.96 | 0.95 |
|---|---|---|---|---|---|---|
| $X_{1\,FP}$ | 6.89 | 6.53 | 6.3 | 6.15 | 6.01 | 5.92 |
| $X_{1\,CFP}$ | 6.89 | 6.04 | 5.6 | 5.29 | 5.06 | 4.89 |
| $(R_*/R_0)_{FP}$ | 1 | 0.981 | 0.963 | 0.946 | 0.930 | 0.918 |
| $(R_*/R_0)_{CFP}$ | 1 | 0.969 | 0.951 | 0.933 | 0.915 | 0.897 |
| $(M_*/M_0)_{FP}$ | 1 | 0.963 | 0.928 | 0.896 | 0.866 | 0.842 |
| $(M_*/M_0)_{CFP}$ | 1 | 0.950 | 0.909 | 0.874 | 0.840 | 0.809 |

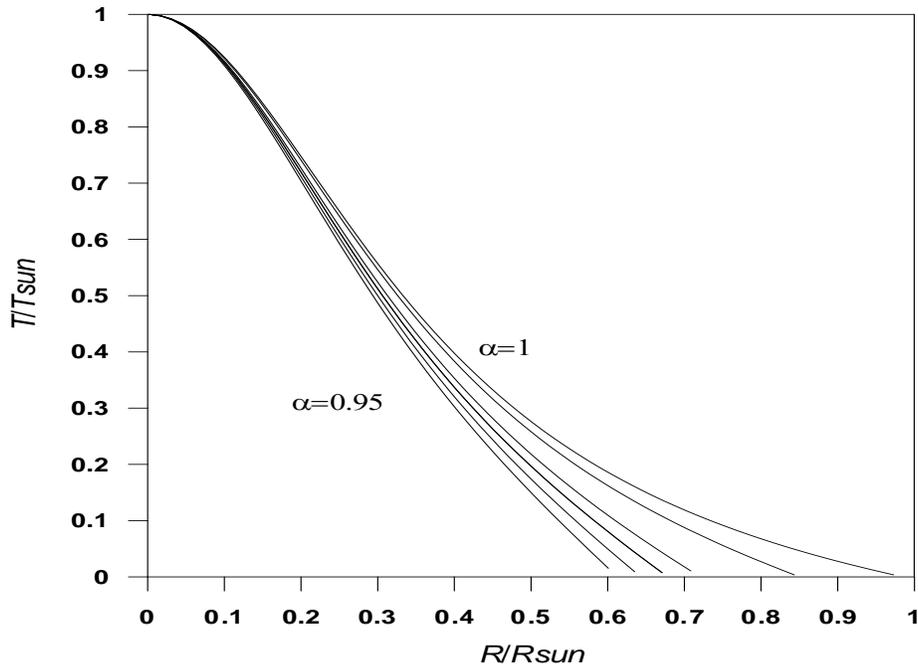

Figure 1. Emden function of the conformable fractional polytrope with n=3.

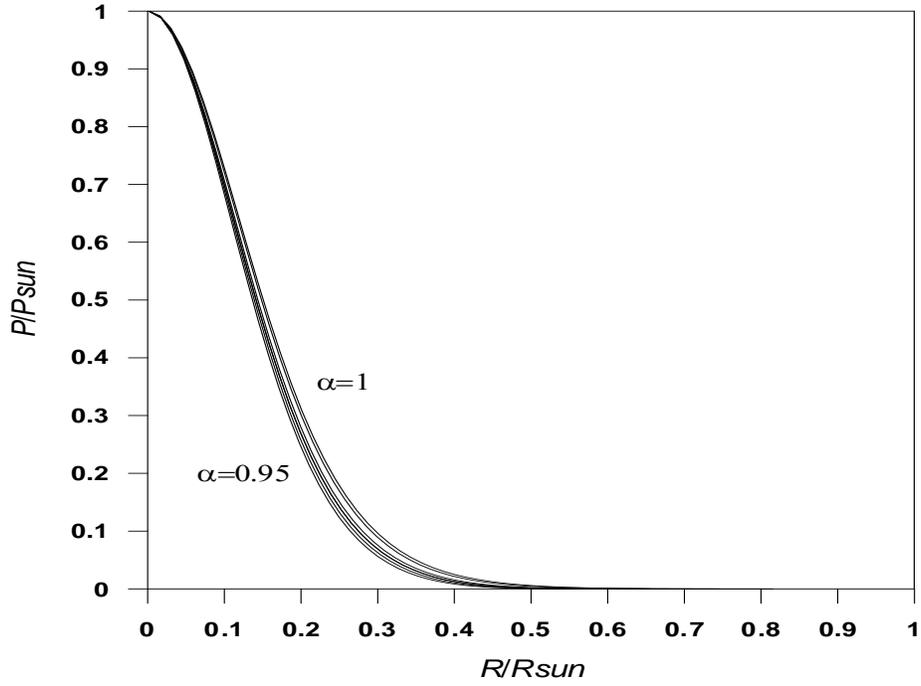

Fig 2: Distribution of the pressure ratio of the conformable fractional polytrope with n=3.

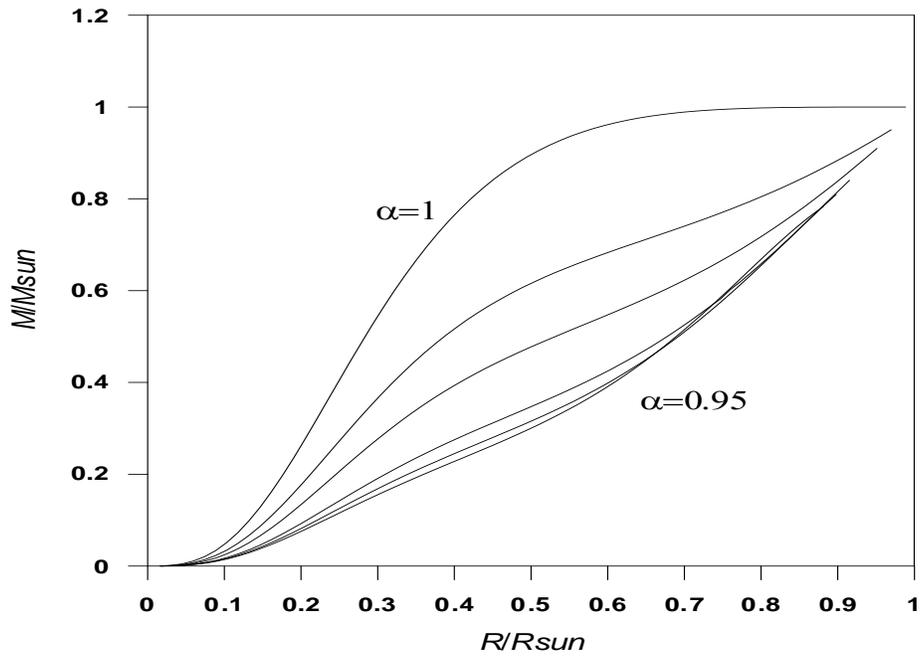

Fig 3: Mass-Radius relation of the conformable fractional polytrope with n=3.

The interior of the compact stars is nearly isothermal and the core temperature approximately equals the temperature at the core envelope-boundary. Because of the pressure of the degenerate matter is nearly independent of the temperature, so polytropic models may be used. White dwarf models use values greater than one. To declare the effects of the fractional parameters on the structure of the white dwarfs, we constructed fractional polytropic models with n=1.5. We assume the ratio of the white dwarf radius to the solar radius $R_{wd}/R_0 = 0.00188$ and the mass ratio is $M_{wd}/M_0 = 1.41096$. The central density is given by Equation (49). Table 3 lists the results, where column 2 is the first zero of the fractional polytrope, column 3 represents the white dwarf radius ratio, column 4 is the white dwarf mass ratio and column 5 is the central density. The effect of the fractional parameter is clear. Figure 4 shows the fractional mass-radius relation for the white dwarf model.

Table 3: White dwarf polytropic models with n=1.5, $R_{wd}/R_0 = 0.00188$, $M_{wd}/M_0 = 1.41096$.

| $\alpha$ | $X_{1\,CFP}$ | $(R_*/R_{wd})$ | $(M_*/M_{wd})$ | $\rho_c$ $(10^9)$ (g/cm$^3$) |
|---|---|---|---|---|
| 1 | 3.653 | 1 | 1 | 1.792 |
| 0.99 | 3.602 | 0.986 | 0.974 | 1.659 |
| 0.98 | 3.552 | 0.974 | 0.950 | 1.544 |
| 0.97 | 3.504 | 0.962 | 0.927 | 1.444 |
| 0.96 | 3.458 | 0.949 | 0.904 | 1.356 |
| 0.95 | 3.412 | 0.939 | 0.884 | 1.277 |

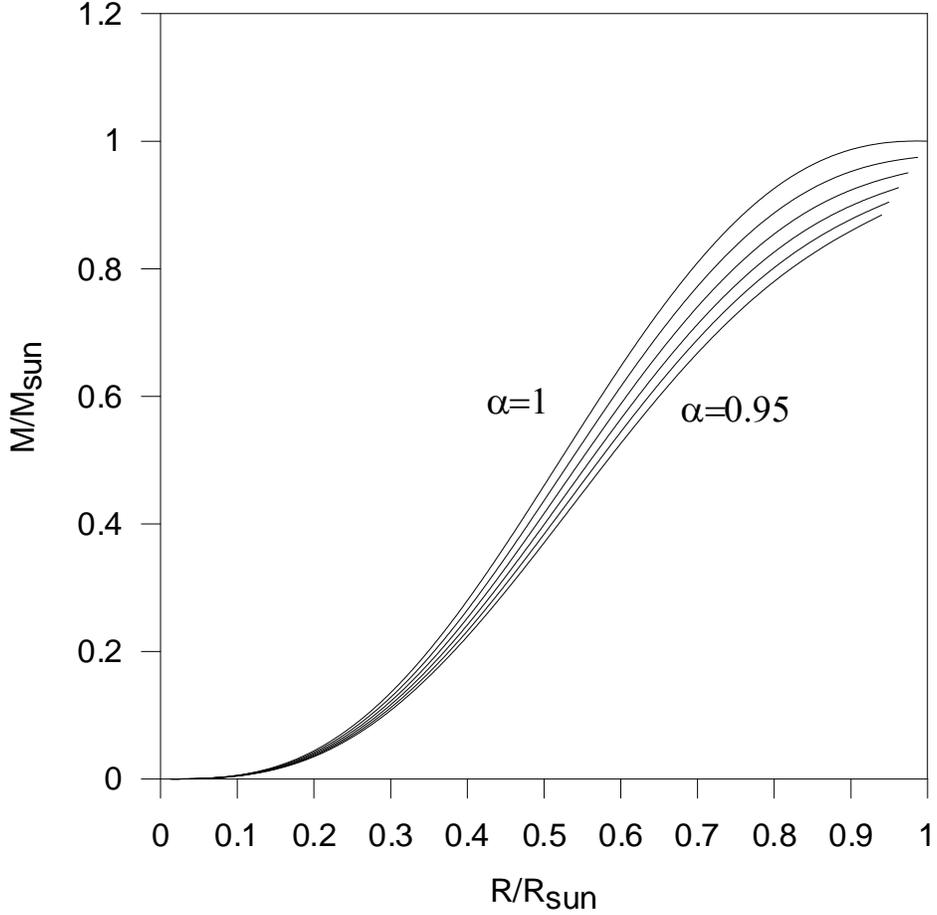

Fig 4: Mass-Radius relation of the conformable fractional polytrope with n=1.5.

## 5. Discussion and Conclusion

To model physical systems via fractional derivatives one need derivative involves an integral over a spatial or time region. In static stellar models, there is no time dependence, then there are no memory effects, but the fractional derivatives of mass, pressure, temperature are given at a particular scaled radius, $r/R$, with $R$ the radius of the star, they do spatially sample the entire star. In the present paper, we solved the fractional Lane-Emden equation using accelerated series expansion. The solution is performed in the frame of conformable fractional derivatives. The calculated models recover the well-known series of solutions when $\alpha = 1$. Physical parameters such as mass-radius relation, density ratio, pressure ratio and temperature ratio for different fractional models have been calculated and investigated. We found that the present models of the conformable fractional stars have smaller volume and mass than that of both the integer star and the fractional models by Nouh and Abd-Elsalam (2018).

Fractional mass-radius relations, pressure distributions and temperature distributions for polytropic indices suitable for the sun and white dwarfs structures were calculated and investigated.